\documentclass[camera]{jpaper}

\usepackage[nocompress]{cite}
\usepackage{algorithmic}
\usepackage{array}

\makeatletter
\let\MYcaption\@makecaption
\makeatother

\usepackage[font=footnotesize]{subcaption}

\makeatletter
\let\@makecaption\MYcaption
\makeatother

\usepackage{fixltx2e}
\usepackage{dblfloatfix}
\usepackage[nolessnomore, italic]{mathastext}
\usepackage[T1]{fontenc}
\usepackage[usenames,dvipsnames,svgnames,table]{xcolor}
\usepackage[normalem]{ulem}
\usepackage{enumitem}
\usepackage{setspace}
\usepackage{indentfirst}
\usepackage{footmisc}
\usepackage{fancyhdr}
\usepackage{authblk}
\usepackage[us,12hr]{datetime}
\usepackage[keeplastbox]{flushend}
\usepackage[hidelinks]{hyperref}

\usepackage{pifont}
\usepackage{xspace}

\newcommand{\titleShort}[0]{MeDiC\xspace}
\newcommand{\titleLong}[0]{Memory Divergence Correction\xspace}
\newcommand*\mycirc[1]
{\tikz[baseline=(char.base)]{
  \node[shape=circle,draw,inner sep=0.5pt,fill=black,text=white,font=\sffamily] (char) {#1};}}

\newboolean{publicversion}
\setboolean{publicversion}{true}

\ifthenelse{\boolean{publicversion}}{
  \pagenumbering{arabic}
  \newcommand{\grumbler}[2]{}
  \newcommand{\assign}[1]{}
  \newcommand{\respond}[3]{}
  \newcommand{\changesI}[0]{}
  \newcommand{\changesII}[0]{}
  \newcommand{\changesIII}[0]{}

}{
  \pagenumbering{arabic}
  \newcommand{\grumbler}[2]{\textcolor{blue}{\bf #1: #2}}
  \newcommand{\assign}[1]{\textcolor{purple}{\bf RESPONSIBLE: #1}}
  \newcommand{\respond}[3]{\textcolor{#1}{\bf #2-response: #3}}
  \newcommand{\changesI}[0]{}
  \newcommand{\changesII}[0]{}
  \newcommand{\changesIII}[1]{\textcolor{BrickRed}{#1}}

}

\newcommand{\red}[1]{#1}
\newcommand{\green}[1]{#1}

\newif\ifcameraready
\camerareadytrue
\newcommand{\versionnum}[0]{5}

\ifcameraready
\else
\fi

\ifcameraready
  \newcommand{\todo}[1][]{}
\else
  \newcommand{\todo}[1][]{\textbf{\fcolorbox{black}{red}{\color{white}{TODO}}} \underline{$\overline{\hbox{\emph{#1}}}$}}
\fi

\begin{document}
\title{Holistic Management of the GPGPU Memory Hierarchy\\to Manage Warp-level Latency Tolerance}

% author names and affiliations
% use a multiple column layout for up to two different
% affiliations

%\author{%
%{Saugata Ghose\affilCMU}%
%\qquad%
%{Yixin Luo\affilCMU}%
%\qquad%
%{Onur Mutlu\affilETH\affilCMU}%
%\vspace{3pt}\\%
%{\it\affilCMU Carnegie Mellon University \qquad \affilETH ETH Z{\"u}rich}%
%\vspace{3pt}%
%}

\author{%
Rachata Ausavarungnirun$^1$
\qquad%
Saugata Ghose$^1$%
\qquad%
Onur Kay{\i}ran$^{2,3}$%
\vspace{2pt}\\%
Gabriel H. Loh$^2$%
\qquad%
Chita R. Das$^3$%
\qquad%
Mahmut T. Kandemir$^3$%
\qquad Onur Mutlu$^{4,1}$}
\affil{
{\it 
    $^1$Carnegie Mellon University\qquad%
    $^2$AMD Research%
\qquad%
    $^3$Pennsylvania State University\qquad%
    $^4$ETH Z{\"u}rich%
}%
}

% make the title area
\maketitle

% \setstretch{0.87}
% \renewcommand{\footnotelayout}{\setstretch{0.85}}

% !TeX root = ../paper.tex

\begin{abstract}

%Graphics Processing Units (GPUs) have enormous parallel processing power that
%can leverage thread-level parallelism. Traditionally, GPUs have been used to
%run graphics applications as they naturally exhibit high amounts of
%concurrency. In recent years, with programming languages such as CUDA and
%OpenCL, programmers have also been able to parallelize non-graphics,
%general-purpose (GPGPU) applications into thousands of threads to harness the
%power of GPUs.

This paper summarizes the idea of Memory Divergence Correction (MeDiC), which was
published at PACT 2015~\cite{medic},
and examines the work's significance and future potential.
In a modern GPU architecture, all threads within a warp execute the same instruction in lockstep.
For a memory instruction, this can lead to \emph{memory divergence}: the memory
requests for some threads are serviced early, while the remaining requests
incur long latencies.  This divergence stalls the warp, as it cannot
execute the next instruction until \emph{all} requests from the current 
instruction complete.

In this work, we make three new observations.  First, GPGPU warps exhibit
heterogeneous memory divergence behavior at the shared cache: some warps have
most of their requests hit in the cache (high cache utility), while other warps
see most of their request miss (low cache utility).  Second, a warp retains the same divergence
behavior for long periods of execution.  Third, due to high memory level
parallelism, requests going to the shared cache can incur queuing delays as 
large as hundreds of cycles, exacerbating the effects of memory divergence.

We propose a set of techniques, collectively called \emph{\titleLong{}}
(\titleShort{}), that reduce the negative performance impact of memory divergence and cache queuing.
\titleShort{} uses online warp divergence characterization to guide three components:
(1)~a cache bypassing mechanism that exploits the latency tolerance of low cache
utility warps to both alleviate queuing delay and increase the hit rate for high
cache utility warps, (2)~a cache insertion policy that prevents data from high
cache utility warps from being prematurely evicted, and (3)~a memory controller
that prioritizes the few requests received from high cache utility warps to
minimize stall time.  We compare \titleShort{} to four cache management 
techniques, and find that it delivers an average speedup of 21.8\%, and 20.1\%
higher energy efficiency, over a state-of-the-art GPU cache management
mechanism across 15 different GPGPU applications.

\end{abstract}

% !TeX root = ../paper.tex

\section{Introduction}
%\section{Memory Divergence in GPUs}
\label{sec:introduction} 

Graphics Processing Units (GPUs) have enormous parallel processing power to
leverage thread-level parallelism.  GPU applications \changesII{are usually} broken down into
thousands of threads, allowing GPUs to use \emph{fine-grained
multithreading}~\cite{cdc6600,smith-hep} to prevent GPU cores from stalling due
to dependencies and long memory latencies.  Ideally, there should always be
available threads for GPU cores to continue execution, preventing stalls within
the core. GPUs also take advantage of the \emph{SIMD} (Single Instruction,
Multiple Data) execution model~\cite{flynn}.  The thousands of threads within a
GPU application are clustered into \emph{thread
blocks}, each of which contains multiple smaller bundles (\emph{warps}) of
threads that run concurrently. \changesII{Each thread in a warp executes the same
instruction on a different piece of data. A warp completes an instruction when all
threads in the warp complete the instruction.}

\begin{figure}[h!]
	\centering
        \vspace{3pt}
	\includegraphics[width=\columnwidth]{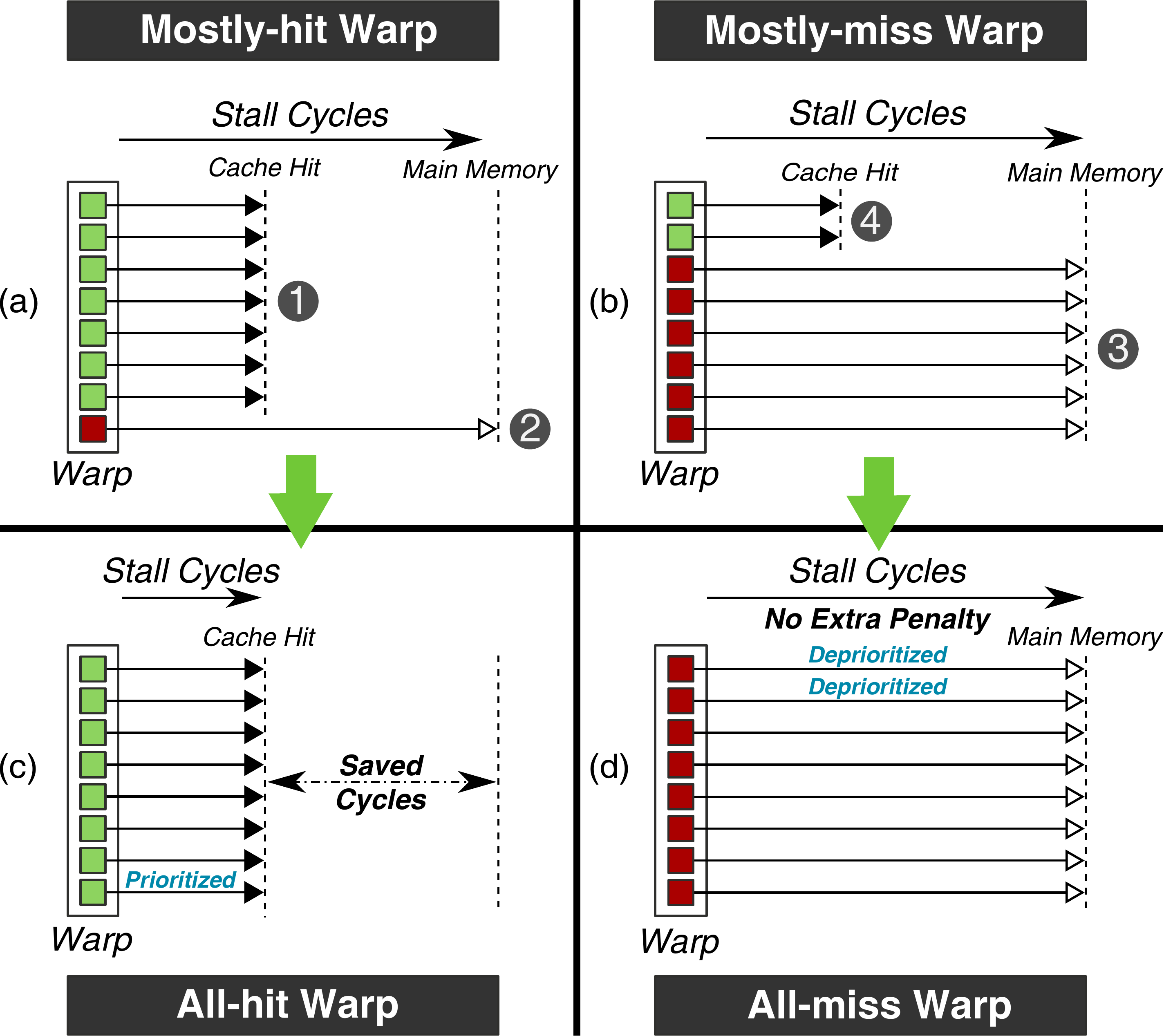}	
    \caption{Memory divergence within a warp. (a) and (b) show the heterogeneity between
    \emph{mostly-hit} and \emph{mostly-miss} warps, respectively.  (c) and (d) show the change in stall time from 
    converting \emph{mostly-hit warps into all-hit warps}, and \emph{mostly-miss warps into all-miss
    warps}, respectively. Reproduced from~\cite{medic}.}
	\label{fig:mem-div}
\end{figure}

While many GPGPU applications can tolerate a significant amount of memory
latency due to their parallelism and the use of fine-grained multithreading,
\emph{memory divergence} (where the threads of a warp reach a memory
instruction, and some of the threads' memory requests take longer to service than others)
can significantly increase the stall time of a warp~\cite{ccws,nmnl-pact13,cpugpu-micro,kuo-throttling14,largewarps,warpsub,zheng-cal,caws-pact14,tor-micro13}.
Because all threads within a warp operate in lockstep due to the SIMD execution
model, the warp cannot proceed to the next instruction until the \emph{slowest}
request within the warp completes. Figures~\ref{fig:mem-div}a and~\ref{fig:mem-div}b show
examples of memory divergence within a warp.
Figure~\ref{fig:mem-div}a shows a \emph{mostly-hit warp}, where most
of the warp's memory accesses hit in the cache (\mycirc{1}). \changesII{Only} a
single access misses in the cache and must go to main memory
(\mycirc{2}). As a result, the \emph{entire warp} is stalled until the
much longer cache miss completes. 
Figure~\ref{fig:mem-div}b shows a \emph{mostly-miss warp}, where 
most of the warp's memory requests miss in the cache (\mycirc{3}), resulting in many
accesses to main memory.  Even though some requests are cache hits (\mycirc{4}),
these do not benefit the execution time of the warp \changesII{since the execution of
the warp ends when the slowest thread finishes the instruction.}

\changesI{
Based on these three observations, we aim to devise a mechanism that has two major goals: (1)~convert mostly-hit warps into
\emph{all-hit warps} (warps where \emph{all} requests hit in the cache, as shown in
Figure~\ref{fig:mem-div}c), and (2)~convert mostly-miss warps into
\emph{all-miss warps} (warps where \emph{none} of the requests hit in the cache,
as shown in Figure~\ref{fig:mem-div}d).  As we can see in
Figure~\ref{fig:mem-div}a, the stall time due to memory divergence for the 
mostly-hit warp can be eliminated by converting only the single cache miss (\mycirc{2}) into
a hit.  Doing so requires additional cache space.  If we convert the two cache
hits of the mostly-miss warp (Figure~\ref{fig:mem-div}b, \mycirc{4}) into cache misses, 
we can \changesII{allocate} the cache space previously used by these hits to the mostly-hit warp, thus converting the
mostly-hit warp into an all-hit warp.  Though the mostly-miss warp is now an
all-miss warp (Figure~\ref{fig:mem-div}d), it incurs no extra stall penalty, as the warp was already waiting on the other six
cache misses to complete. \changesII{Moreover}, now that it is an all-miss warp, 
we \changesII{can} predict that its future memory requests will also not be in the L2 cache. \changesII{Based on this prediction,}
we can simply have these
requests \emph{bypass the cache}. \changesII{By} doing so, the requests from the all-miss
warp can completely avoid unnecessary L2 access and queuing delays, \changesII{and enable the use of 
L2 cache bandwidth and buffer space by warps that benefit from the L2 cache}.  This decreases the total
number of requests going to the L2 cache, thus reducing the queuing latencies for
requests from mostly-hit and all-hit warps, as there is less contention.
}

\section{Observation on GPU Memory Divergence}

\changesI{
We make three new key observations about memory divergence (at the shared L2 cache).
First, we observe that the degree of memory divergence can differ across warps \changesII{(as illustrated in Figure~\ref{fig:mem-div})}. 
This inter-warp heterogeneity affects how well each warp takes advantage of
the shared cache.  Second, we observe that a warp's memory divergence behavior tends
to remain stable for long periods of execution, making it
predictable.  Third, we observe that requests to the shared cache experience
long queuing delays due to the large amount of parallelism in GPGPU programs,
which exacerbates the memory divergence problem and slows down GPU execution.
Next, we describe each of these observations in detail and motivate our solutions.}

\subsection{Memory Divergence Heterogeneity}
\label{sec:warp-type}

%\vspace{3pt}
There is \emph{heterogeneity across warps} in
the degree of memory divergence experienced by each warp at the shared L2
cache. \changesII{Figures~\ref{fig:mem-div}a and \ref{fig:mem-div}b show} examples of two different \emph{types of
warps} that exhibit different degrees of memory divergence.
%Figure~\ref{fig:mem-div}a shows a \emph{mostly-hit warp}, where most of the
%warp's memory accesses hit in the cache (\mycirc{1}). However, a single access
%misses in the cache and must go to main memory (\mycirc{2}) and the
%\emph{entire warp} is stalled until the much longer cache miss completes.
%Figure~\ref{fig:mem-div}b shows a \emph{mostly-miss warp}, where most of the
%warp's memory requests miss in the cache (\mycirc{3}), resulting in many
%accesses to main memory.  Even though some requests are cache hits
%(\mycirc{4}), these do not benefit the execution time of the warp.

\begin{figure}[h!]
     \vspace{3pt}
	\centering
	\includegraphics[width=\columnwidth]{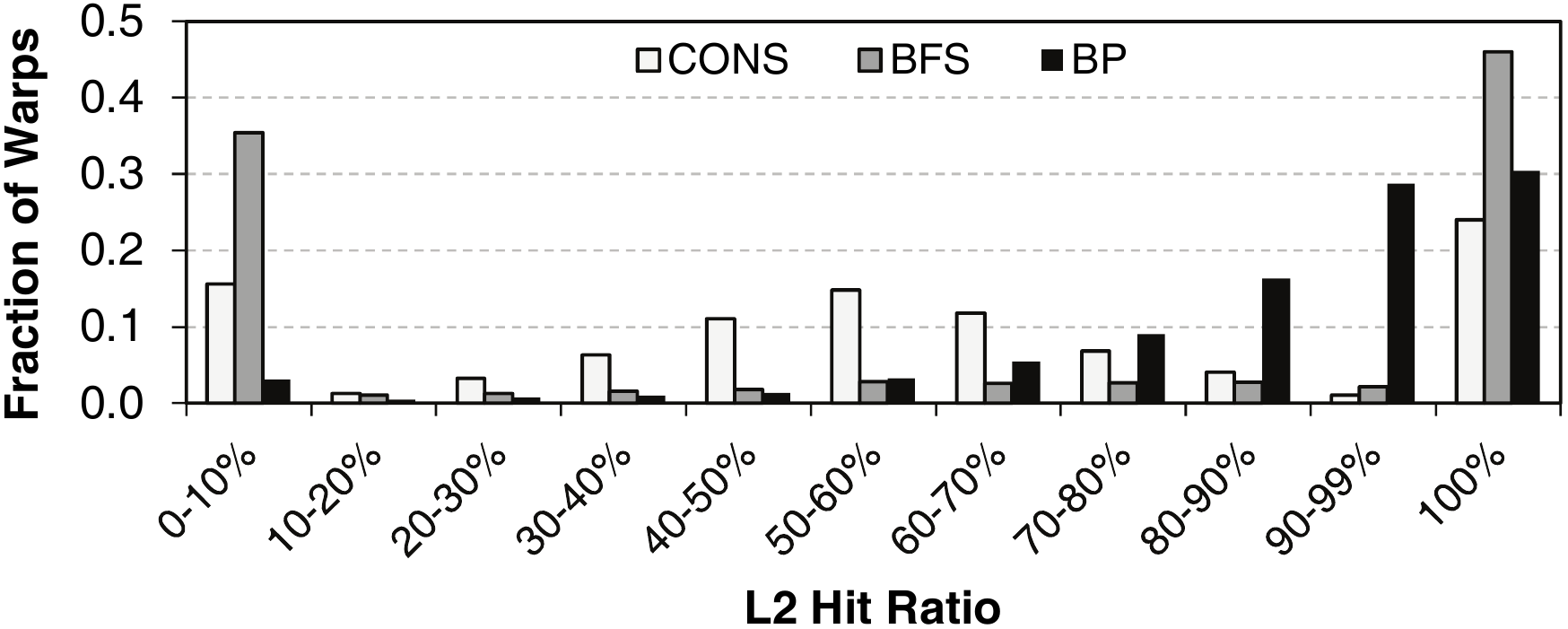}%
	\caption{L2 cache hit ratio of different warps in three representative GPGPU applications. Reproduced from~\cite{medic}.} 
        %\vspace{-10pt}
	\label{fig:warp-spread}
\end{figure}

\begin{figure}[h!]
        %\vspace{-5pt}
	\centering
	\includegraphics[width=0.4\textwidth]{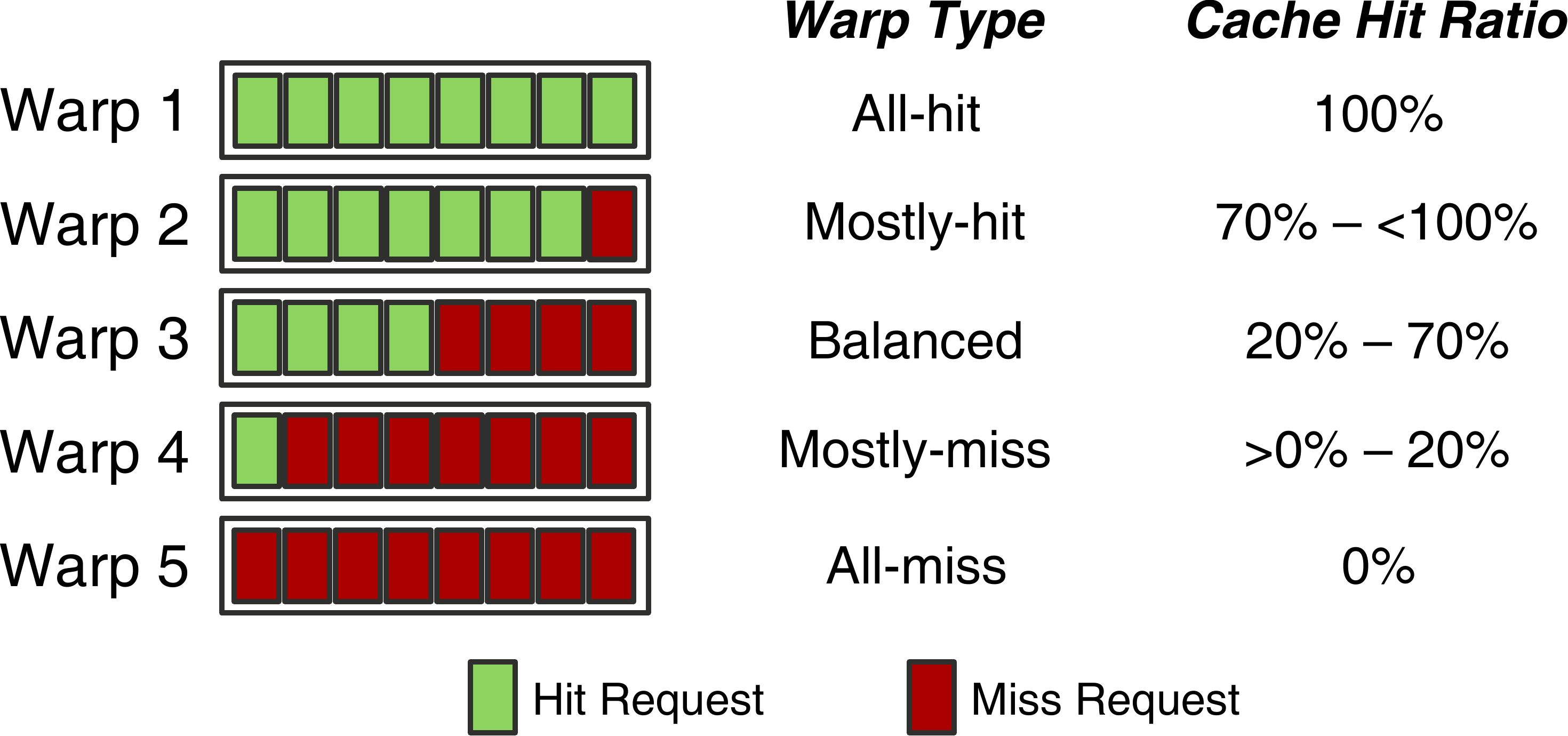}%
	%\vspace{-1pt}
	\caption{Warp type categorization based on the shared cache hit ratios.
    Hit ratio values are empirically chosen. Reproduced from~\cite{medic}.} 
	\label{fig:warp-types}
\end{figure}

We observe that different warps have different amounts of sensitivity to memory
latency and cache utilization.  We study the cache utilization of a warp by
determining its \emph{hit ratio}, the percentage of memory requests that hit in
the cache when the warp issues a single memory instruction. As
Figure~\ref{fig:warp-spread} shows, the warps from each of our three
representative GPGPU applications are distributed across all possible ranges of
\emph{hit ratio}, exhibiting significant heterogeneity.  To better characterize
warp behavior, we break the warps down into the five types shown in
Figure~\ref{fig:warp-types} based on their hit ratios: \emph{all-hit},
\emph{mostly-hit}, \emph{balanced}, \emph{mostly-miss}, and \emph{all-miss}.

MeDiC provide two mechanisms, warp-type-aware cache bypassing and
warp-type-aware cache insertion policy, \changesII{in order} to convert \emph{mostly-hit} warps
into \emph{all-hit} warps, where all requests in the warp hit in the cache,
\changesII{thereby} reducing the stall time of mostly-hit warp significantly. This is done at
the cost of converting the \emph{mostly-miss} warps into \emph{all-miss}
warps, \changesII{but doing so does not increase the stall time of such warps}.
To speed up uncacheable cache misses from mostly-hit warps, the
warp-type-aware memory scheduling policy in \titleShort prioritizes memory
requests from mostly-hit warps over memory requests from mostly-miss warps.

\subsection{Memory Divergence Stability Over Time}
\begin{figure}[h!]
	\centering
	%\vspace{-5pt}
	\includegraphics[width=\columnwidth]{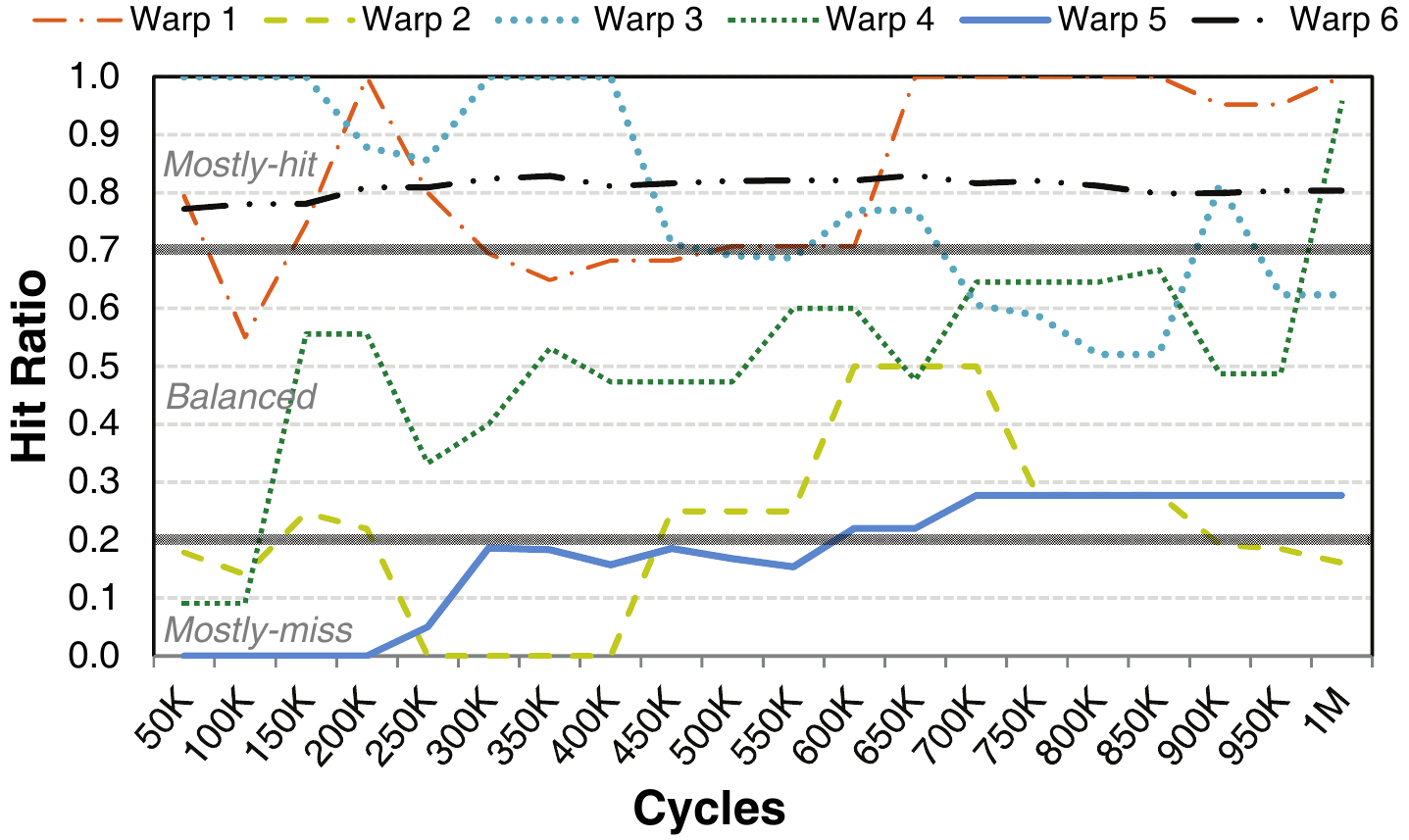}%	
	\caption{Hit ratio of randomly selected warps over time \changesII{from BFS}. Reproduced from~\cite{medic}.} 
	\label{fig:warp-ratio-trend}
\end{figure}

\emph{A warp tends to retain its
memory divergence behavior} (e.g., whether or not it is mostly-hit or
mostly-miss) \emph{for long periods of execution}, and is thus predictable.
This is due to the spatial and temporal locality of each thread within the
warp. Figure~\ref{fig:warp-ratio-trend} shows a sample of warps from a \changesII{representative}
application (i.e., BFS~\cite{lonestar}) that shows this predictability. This predictability enables us to
perform history-based warp divergence characterization. 

%In MeDiC, we include warp type identification logic, which profiles and categorize warps 
%into one of the five categories: \emph{all-hit},
%\emph{mostly-hit}, \emph{balanced}, \emph{mostly-miss} and \emph{all-miss}. To
%cope with long-term shifts in warp divergence behavior, 
%he hit ratio is periodically resampled, and the warp type is adjusted if need be.
%
%
%\begin{figure}[t]
%        %\vspace{-5pt}
%	\centering
%	\includegraphics[width=0.4\textwidth]{figs/four-category.pdf}%
%	%\vspace{-1pt}
%	\caption{Warp type categorization based on the shared cache hit ratios.
%    Hit ratio values are empirically chosen.} 
%	\label{fig:warp-types}
%\end{figure}

% We also show that a warp tends to retain their memory divergence behavior
% throughout its lifetime. This predictability enables us to perform history-based warp
% characterization as we will show in Section~\ref{sec:mech}.

%Second, we observed that the same warp tends to exhibit the similar memory
%divergence pattern. In particular, we observe that warps that have most of its
%threads miss at the cache tend to retain its divergence characteristics. This
%allows an easy method to keep track of the divergence in each warp. In this
%paper, we take advantage of this observation and use a simple per-warp counter
%to keep track of the divergence information. Additionally, we proposed several
%improvements to how memory requests is handled at the cache and main memory
%in order to take advantage of this divergence information.

\subsection{High Queuing Latencies at the Shared Cache}
\begin{figure}[h!]
	\centering
	\includegraphics[width=\columnwidth]{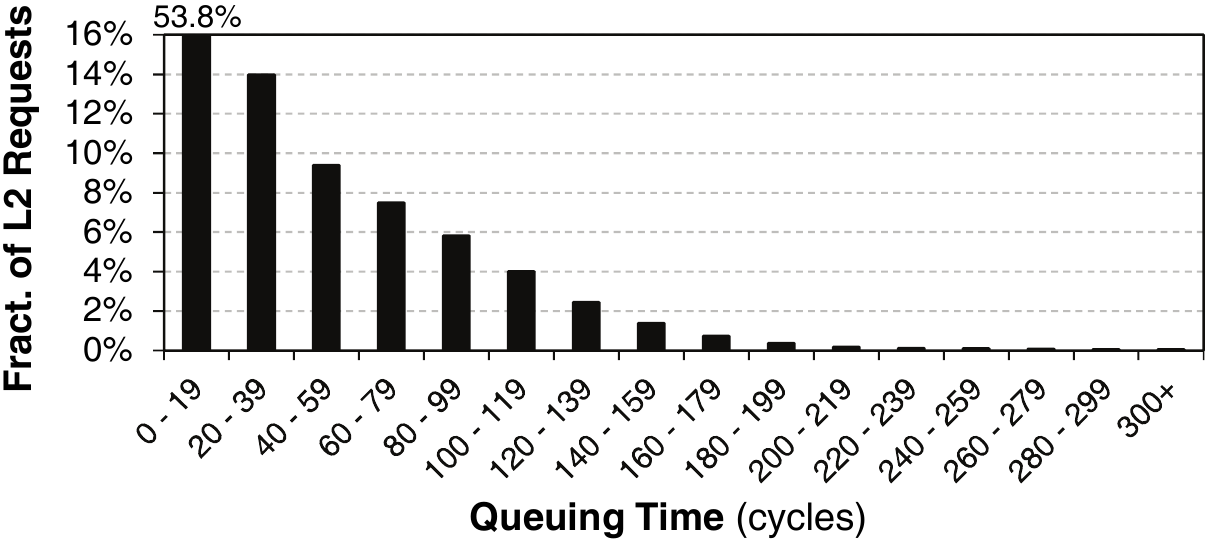}	
    \caption{Distribution of per-request queuing latencies for L2 cache requests from BFS. Reproduced from~\cite{medic}.}
	\label{fig:l2-queue-latency}
\end{figure}

Due to the amount of thread parallelism
within a GPU, \emph{a large number of memory requests can arrive at the L2
cache in a small window of execution time, leading to significant queuing
delays}.  Prior work observes high access
latencies for the shared L2 cache within a GPU~\cite{sisoftware,
coherence-hpca13,demystify}, but does not identify \emph{why}
these latencies are so high.  We show that when a large number of requests arrive
at the L2 cache, both the limited number of read/write ports and backpressure
from cache bank conflicts force many of these requests to queue up for long
periods of time. We observe that this queuing latency can sometimes add
\emph{hundreds} of cycles to the cache access latency, and that non-uniform
queuing across the different cache banks exacerbates memory divergence. Figure~\ref{fig:l2-queue-latency} quantifies the magnitude of this queue contention \changesII{if we set} the cache lookup
latency at one cycle, \changesII{for one application, BFS~\cite{lonestar}}. As shown, a significant number of requests experience tens
to hundreds of cycles of queuing delay.

The warp-type-aware bypassing logic in MeDiC helps to alleviate these L2
queuing latencies.  By preventing mostly-miss and all-miss warps from
accessing the cache, which yields little \changesII{or no} benefit to them, we reduce
the access latencies for requests from \changesII{(1)}~mostly-hit and all-hit warps, \changesII{which
benefit from the cache much more; and also (2)~mostly-miss and all-miss warps themselves; thereby improving the
overall performance of all warps and the system.}

\begin{figure*}[t!]
	\centering
	\includegraphics[width=\textwidth]{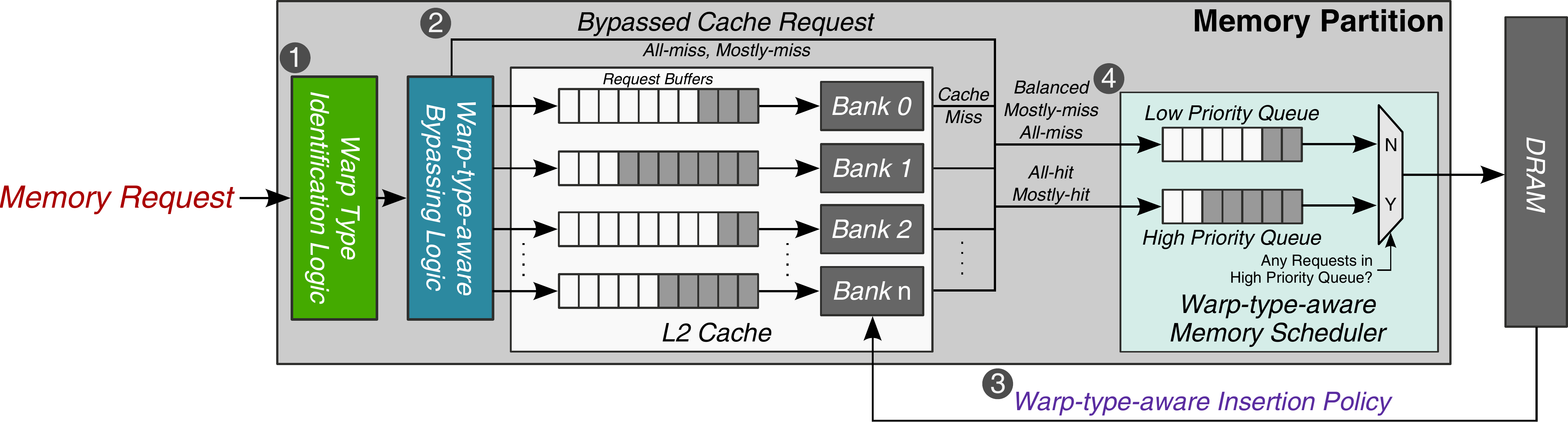}
	\vspace{2pt}
	\caption{Overview of \titleShort{}: \protect\mycirc{1} warp type identification logic, \protect\mycirc{2} warp-type-aware cache bypassing, \protect\mycirc{3} warp-type-aware cache insertion policy, \protect\mycirc{4} warp-type-aware memory scheduler. Reproduced from~\cite{medic}.} 
	\label{fig:main-mechanism}
	\vspace{10pt}
\end{figure*}

%\subsection{Impact of MeDiC}
\section{\titleShort{}: \titleLong}
Based on these three \changesII{new} observations \changesII{we made, we define three} %and the key ideas of \titleShort, 
major goals \changesII{for our new mechanism. We would like to devise a mechanism that} (1)~converts mostly-hit warps into \emph{all-hit warps}
(warps where \emph{all} requests hit in the cache, as shown in
Figure~\ref{fig:mem-div}c), (2)~converts mostly-miss warps into \emph{all-miss
warps} (warps where \emph{none} of the requests hit in the cache, as shown in
Figure~\ref{fig:mem-div}d) and (3)~reduces L2 cache queuing delay for all warp
types. As we can see in Figure~\ref{fig:mem-div}a, the stall time due to memory
divergence for the mostly-hit warp can be eliminated by converting only the
single cache miss (Figure~\ref{fig:mem-div}a, \mycirc{2}) into a \changesII{cache} hit. 

\changesI{To this end,} we introduce \emph{\titleLong{}}
(\emph{\titleShort{}}), a GPU-specific mechanism that exploits \emph{memory
divergence heterogeneity} across warps at the shared cache and at main memory
to improve the overall performance of GPGPU applications.  \titleShort{}
consists of three different components, which work together to achieve our \changesII{three}
goals: % of converting mostly-hit warps into all-hit warps and mostly-miss warps into all-miss warps: 
(1)~a warp-type-aware \emph{cache bypassing mechanism},
which prevents requests from mostly-miss and all-miss warps from accessing the
shared L2 cache; (2)~a warp-type-aware \emph{cache insertion policy}, which
prioritizes requests from mostly-hit and all-hit warps, \changesII{in order to increase the likelihood} that they all
become cache hits; and (3)~a warp-type-aware \emph{memory scheduling
mechanism}, which prioritizes requests from mostly-hit warps that were not
successfully converted to all-hit warps, in order to minimize the stall time
due to divergence. These three components are all driven by an online mechanism
that can identify the expected memory divergence behavior of each warp.

%\subsection{Putting Everything Together: \titleLong}

Figure~\ref{fig:main-mechanism} shows the overall \titleShort{} mechanism.
\titleShort{} consists of \changesII{four} different components: \mycirc{1} a
\emph{warp-type-identification mechanism} that classifies warps into one of the four
warp types as described in Section~\ref{sec:warp-type};
\mycirc{2} a \emph{bypass
mechanism} that bypasses requests from all-miss and mostly-miss warps, reducing
the queuing delay in the L2 cache; \mycirc{3} an \emph{insertion policy} that prevent mostly-hit
requests from being evicted from the cache; and \mycirc{4} a \emph{memory scheduler}
that prioritizes requests from mostly-hit warps, which are more latency
sensitive.

\subsection{Warp Type Identification}
\label{sec:identify}

In order to take advantage of the memory divergence heterogeneity across warps,
we must first add hardware that can identify the divergence behavior of each
warp.  The key idea is to periodically sample the hit ratio of a warp, and to
classify the warp's divergence behavior as one of the five types in Figure~\ref{fig:warp-types} 
based on the observed hit ratio. This information can then be used to drive the
warp-type-aware components of \titleShort{}.  In general, warps tend to retain
the same memory divergence behavior for long periods of execution. However,
there can be some long-term shifts
in warp divergence behavior, requiring periodic resampling of the hit ratio to
potentially \changesII{re-evaluate} the warp type.
Warp type identification through hit ratio sampling
requires hardware within the cache to periodically count the number of hits
and misses each warp incurs.  We append two counters to the metadata stored
for each warp, which represent the total number of cache hits and
cache accesses for the warp \changesII{during the sampling interval}.

\subsection{Warp-type-aware Shared Cache Bypassing}
\label{sec:bypass}

Once the warp type is known and a warp generates a request to the L2 cache, our
mechanism first decides whether to bypass the cache based on the warp type.
The key idea behind \emph{warp-type-aware cache bypassing} is to convert
mostly-miss warps into all-miss warps, as they do not benefit greatly from the
few cache hits that they get.  By bypassing these requests, we achieve three
benefits: (1)~bypassed requests can avoid L2 queuing latencies entirely,
(2)~other requests that do hit in the L2 cache experience shorter queuing
delays due to the reduced contention, and (3)~space is created in the L2 cache
for mostly-hit warps.

The cache bypassing logic must make a simple decision: if an incoming memory
request was generated by a mostly-miss or all-miss warp, the request is bypassed
directly to DRAM.  This is determined by reading the warp type stored in the
warp metadata from the warp type identification mechanism.  A simple 2-bit
demultiplexer can be used to determine whether a request is sent to the L2 bank
arbiter, or directly to the DRAM request queue.

\subsection{Warp-type-aware Cache Insertion Policy}
\label{sec:insertion}

Our cache bypassing mechanism frees up space within the L2 cache, which we want
to use for the cache misses from mostly-hit warps (to convert \changesII{the cache miss} memory
requests into cache hits). However, even with the new bypassing mechanism,
other warps (e.g., balanced, mostly-miss) still insert some data into the
cache. In order to aid the conversion of mostly-hit warps into all-hit warps,
we develop a \emph{warp-type-aware cache insertion policy}, whose key idea is
to ensure that \changesII{in} a given cache set, data from mostly-miss warps are evicted
first, while data from mostly-hit warps and all-hit warps are evicted last.

To ensure that a cache block from a mostly-hit warp stays in the cache for as
long as possible, we insert the block closer to the MRU position.  A cache
block requested by a mostly-miss warp is inserted closer to the LRU position,
making it more likely to be evicted.  To track the \changesII{warp type associated with} these cache
blocks, we add two bits of metadata to each cache block, indicating the warp
type. These bits are then appended to the replacement
policy bits. The bits modify the replacement policy behavior, such that a cache block from a mostly-miss warp is more likely
to get evicted than a block from a balanced warp.  Similarly, a cache block
from a balanced warp is more likely to be evicted than a block from a
mostly-hit or all-hit warp.

\subsection{Warp-type-aware Memory Scheduler}
\label{sec:mem-controller}

%\begin{figure}[t!]
%	\centering
%	\includegraphics[width=0.45\textwidth]{results/warp_types.pdf}	
%	\caption{A breakdown of different warp categories on GPGPU applications we evaluated.} 
%	\label{fig:warp-type-gpgpu}
%\end{figure}

Our cache bypassing mechanism and cache insertion policy work to increase the
likelihood that \emph{all} requests from a mostly-hit warp become cache hits,
converting the warp into an all-hit warp.  However, due to cache conflicts, or
due to poor locality, there may still be cases when a mostly-hit warp cannot be
fully converted into an all-hit warp, and is therefore unable to avoid stalling
due to memory divergence as at least one of its requests has to go to DRAM. In
such a case, we want to minimize the amount of time that this warp stalls. To
this end, we propose a \emph{warp-type-aware memory scheduler} that prioritizes
the occasional DRAM requests from mostly-hit warps.

The design of our memory scheduler is very simple. Each memory request is
tagged with a single bit, which is set if the memory request comes from a
mostly-hit warp (or an all-hit warp, in case the warp was mischaracterized).
We modify the request queue at the memory controller to contain two different
queues, where a \emph{high-priority queue} contains all requests that have
their mostly-hit bit set to one.  The \emph{low-priority queue} contains all
other requests, whose mostly-hit bits are set to zero. Each queue uses
FR-FCFS~\cite{fr-fcfs,frfcfs-patent} as the scheduling policy; however, the
scheduler always selects requests from the high priority queue over requests in
the low priority queue.\footnote{Using two queues ensures that high-priority
requests are not blocked by low-priority requests even when the low-priority
queue is full. Two-queue priority also uses simpler logic design than
comparator-based priority~\cite{bliss,bliss-tpds,sms}.}

\changesII{We describe each component of \titleShort in more detail in Sections 4.1, 4.2,
4.3 and 4.4 of our PACT 2015 paper~\cite{medic}.}

%
%Doing so
%requires additional cache space. With the warp-type-aware cache bypassing and
%warp-type-aware cache insertion policy of \titleShort. We can convert the two
%cache hits of the mostly-miss warp (Figure~\ref{fig:mem-div}b, \mycirc{4}) into
%cache misses, we can cede the cache space previously used by these hits to the
%mostly-hit warp, thus converting the mostly-hit warp into an all-hit warp.
%Though the mostly-miss warp is now an all-miss warp
%(Figure~\ref{fig:mem-div}d), it incurs no extra stall penalty, as the warp was
%already waiting on the other six cache misses to complete. 

% Additionally, now
% that it is an all-miss warp, we predict that its future memory requests will
% also not be in the L2 cache, so we can simply have these requests \emph{bypass
% the cache}. In doing so, the requests from the all-miss warp can completely
% avoid unnecessary L2 access and queuing delays. This decreases the total number
% of requests going to the L2 cache, thus reducing the queuing latencies for
% requests from mostly-hit and all-hit warps, as there is less contention.

\begin{figure*}[ht]
	\centering
	\includegraphics[width=\textwidth]{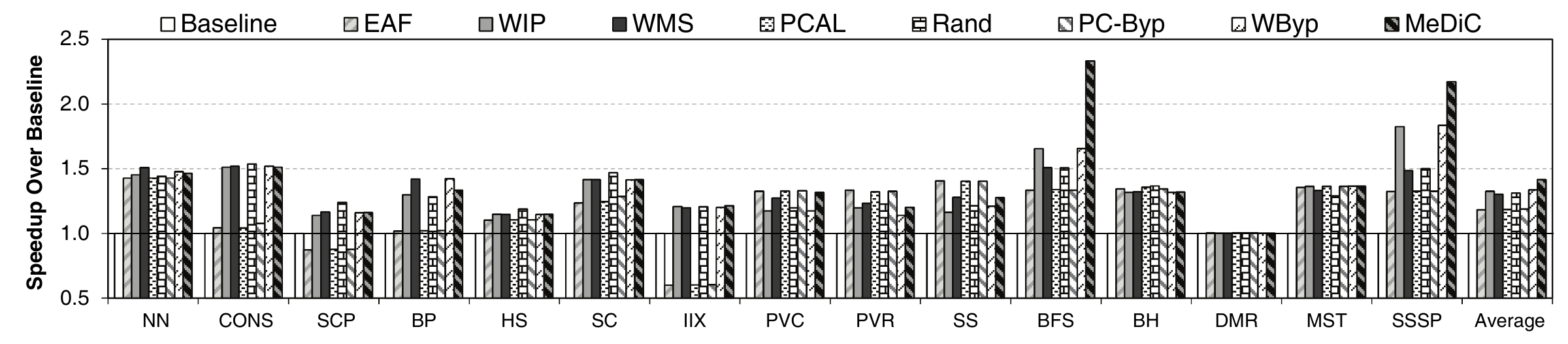}%	
	\caption{\red{Performance of \titleShort{}.} Adapted from~\cite{medic}.} 
	\label{fig:main-res}
	\vspace{5pt}
\end{figure*}

\begin{figure*}[ht]
	\centering
	\includegraphics[width=\textwidth]{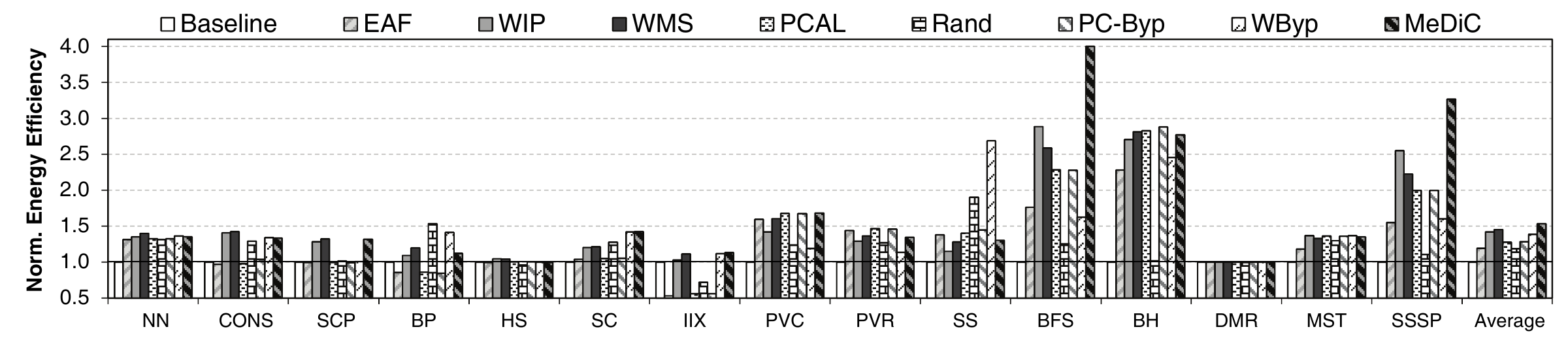}%	
	\caption{\red{Energy efficiency of \titleShort{}.} Adapted from~\cite{medic}.}
	\label{fig:energy-eff}
%	\vspace{-6pt}
\end{figure*}

\section{Methodology}

We model our mechanism using GPGPU-Sim 3.2.1~\cite{gpgpu-sim}.  We modified
GPGPU-Sim to accurately model cache bank conflicts, and added the cache
bypassing, cache insertion, and memory scheduling mechanisms needed to support
\titleShort{}. We use GPUWattch~\cite{gpuwattch} to evaluate power consumption.
\changesIII{We have open sourced our simulator source code at~\cite{medic-sim}.}
We evaluate our system across multiple
GPGPU applications from the CUDA SDK~\cite{cuda-sdk}, Rodinia~\cite{rodinia},
MARS~\cite{mars}, and Lonestar~\cite{lonestar} benchmark suites.%These
%applications are listed in Table~\ref{table:apps}, along with the breakdown of
%warp characterization. 
%The dominant warp type for each application is marked in
%\emph{bold} (AH:~all-hit, MH:~mostly-hit, BL:~balanced, MM:~mostly-miss, 
%AM:~all-miss; see Figure~\ref{fig:warp-types}). 
%We simulate 500 million
%instructions for each kernel of our application, though some kernels complete
%before reaching this instruction count. 

We report performance results using the
harmonic average of the IPC speedup (over the baseline GPU) of each kernel of
each application.\footnote{We confirm that for each application, all kernels
have similar speedup values, and that aside from SS and PVC, there are no
outliers (i.e., no kernel has a much higher speedup than the other kernels).
To verify that harmonic speedup is not swayed greatly by these few outliers, we
recompute it for SS and PVC \emph{without} these outliers, and find that the
outlier-free speedup is within 1\% of the harmonic speedup we report in the
paper.}  Harmonic speedup~\cite{harmonic_speedup,luo-ispass2001} was shown to reflect the average normalized execution
time in multiprogrammed workloads. We calculate energy
efficiency for each workload by dividing the IPC by the energy consumed \changesII{Section 5 of our PACT
2015 paper~\cite{medic} provides more detail on our experimental methodology}.

\section{Evaluation}

\changesI{
Figure~\ref{fig:main-res} shows the performance of \titleShort{} compared to
four GPU cache management mechanisms: the Evicted Address Filter insertion
policy~\cite{eaf-vivek} (\textbf{EAF}), PCAL bypassing policy~\cite{donglihpca15} (\textbf{PCAL}), PC-based
cache bypassing policy (\textbf{PC-Byp}) and \changesII{an idealized} random bypassing policy (\textbf{Rand}) over 15 different GPGPU
applications from 4 benchmark suites. 
We also show the performance of each individual component of \titleShort: our warp-type-aware 
insertion policy (\textbf{WIP}), our warp-type-aware memory scheduling policy (\textbf{WMS}) 
and our warp-type-aware bypassing policy (\textbf{WByp}).

We found that each component of \titleShort{} individually provides significant
performance improvement: WIP (32.5\%), WMS (30.2\%), and WByp (33.6\%).
\titleShort{}, which combines all three mechanisms, provides a 41.5\% performance
improvement over Baseline, on average.  \titleShort{} matches or outperforms its
individual components for all benchmarks except BP, where \titleShort{} has a 
higher L2 miss rate and lower row buffer locality than WMS and WByp.

Our insertion policy, WIP, outperforms EAF~\cite{eaf-vivek} by
12.2\%. We observe that the key benefit of WIP is that cache blocks from
mostly-miss warps are much more likely to be evicted. In addition, WIP reduces
the cache miss rate of several applications. 
Our memory scheduler, WMS, provides significant performance gains
(30.2\%) over Baseline, because the memory scheduler prioritizes requests from
warps that have a high hit ratio, allowing these warps to become active much
sooner than they do in Baseline. 
Our bypassing mechanism, WByp provides an average 33.6\% performance
improvement over Baseline, because it is effective at reducing the L2 queuing
latency..

Compared to PCAL~\cite{donglihpca15}, WByp provides 12.8\% better performance, and
full \titleShort{} provides 21.8\% better performance. We observe that
while PCAL reduces the amount of cache thrashing, the reduction in thrashing
does not directly translate into better performance. We observe that
warps in the mostly-miss category sometimes have high reuse, and acquire tokens to
access the cache. This causes less cache space to become available for mostly-hit warps,
limiting how many of these warps become all-hit. However, when high-reuse warps
that possess tokens are mainly in the mostly-hit category (PVC, PVR, SS, and BH), we find that
PCAL performs better than WByp.

Compared to Rand,\footnote{\green{Note that our evaluation uses an ideal random
bypassing mechanism, where we manually select the best individual
percentage of requests to bypass the cache for each workload.  As a result, the
performance shown for Rand is better than can be practically realized.}} \titleShort{} performs
6.8\% better, because \titleShort{} is able to make bypassing decisions
that do not increase the miss rate significantly. This leads to lower off-chip
bandwidth usage under \titleShort{} than under Rand. 
Rand increases the cache miss rate by 10\% for the kernels of
several applications (BP, PVC, PVR, BFS, and MST). We observe that in many
cases, \titleShort{} improves the performance of applications that tend to
generate a large number of memory requests, and thus experience substantial
queuing latencies. 

Compared to PC-Byp, \titleShort performs 12.4\% better. We observe
that the overhead of tracking the PC becomes significant, and that thrashing
occurs as two PCs can hash to the same index, leading to inaccuracies in the
bypassing decisions.

We conclude that each component of \titleShort{}, and the full
\titleShort{} framework, are effective. Note that each component of
\titleShort{} addresses the same problem (i.e., memory divergence of threads
within a warp) using different techniques on different parts of the memory
hierarchy. For the majority of workloads, one optimization is enough. However,
we see that for certain high-intensity workloads (BFS and SSSP), the congestion is
so high that we need to attack divergence on multiple fronts. Thus,
\titleShort{} provides better average performance than all of its individual
components, especially for such memory-intensive workloads.

We provide the following \changesII{other evaluation} results in our PACT 2015 paper~\cite{medic}:

\begin{itemize}
\item Impact of \titleShort on cache miss rate.
\item Impact of \titleShort on queuing latency.
\item Impact of \titleShort on row buffer locality.
\item Analysis of reuse in GPGPU applications.
\item Hardware cost of \titleShort.
\end{itemize}

\section{Related Work}
\label{sec:related}
%There are several previous works that propose cache
%bypassing as their key mechanism. In this subsection, we will discuss and
%compare \titleShort{} with these proposals. 

To our knowledge, \titleShort{} is the first work that identifies inter-warp memory divergence
heterogeneity and exploits it to achieve better system performance in GPGPU
applications. Our new mechanism consists of warp-type-aware components for
cache bypassing, cache insertion, and memory scheduling. We have already
provided extensive quantitative and qualitative comparisons to state-of-the-art
mechanisms in GPU cache bypassing~\cite{donglihpca15}, cache
insertion~\cite{eaf-vivek}, and memory scheduling~\cite{fr-fcfs,frfcfs-patent}. In
this section, we discuss other related work in these areas.

\textbf{Hardware-based Cache Bypassing.} \green{PCAL is a bypassing mechanism that
addresses the cache thrashing problem by throttling the
number of threads that time-share the cache at any given time~\cite{donglihpca15}. The key idea of
PCAL is to limit the number of threads that get to access the cache. 
%While this work and PCAL share the same goal of reduce cache thrashing, we directly
%address the bigger problem of increasing the utilization of shared resources of
%different warps. Additionally, this work also provides a mechanism that
%differentiates how each warp utilizes its shared resources and provide a
%prioritization scheme at various places, not only in the cache, to minimize the
%amount of time warps are stalled due to long latency memory instructions.
Concurrent work by Li et al.~\cite{li-ics15} proposes a cache bypassing
mechanism that allows only threads with high reuse to utilize the cache. The
key idea is to use locality filtering based on the reuse characteristics of GPGPU
applications, with only high reuse threads having access to the cache.
Xie et al.~\cite{xie-hpca15} propose a bypassing mechanism at the
thread block level. In their mechanism, the compiler statically marks whether thread blocks prefer caching or bypassing. 
At runtime, the mechanism dynamically selects a subset of thread blocks to use
the cache, to increase cache utilization.

Chen et al.~\cite{chen-micro47,chen-mes14} propose a combined warp throttling and
bypassing mechanism for the L1 cache based on the cache-conscious warp
scheduler~\cite{ccws}. The key idea is to bypass the cache when resource
contention is detected. This is done by embedding history information into the
L2 tag arrays. The L1 cache uses this information to perform bypassing
decisions, and only warps with high reuse are allowed to access the L1 cache. 
Jia et al. propose an L1 bypassing mechanism~\cite{mrpb}, whose
key idea is to bypass requests when there is an associativity stall.
Dai et al. propose a mechanism to bypass cache based on a model of a cache
miss rate~\cite{dai-dac16}.
%As we have shown in Section~\ref{sec:eval}, our mechanism is not only able to identify highly reused
%blocks implicitly, but our mechanism can also differentiate between warps with high reuse that
%are likely to benefit from the cache and warps with high reuse that are unable to benefit from
%the cache.} 

\titleShort{} differs from these prior cache bypassing works because it
uses warp memory divergence heterogeneity for bypassing decisions. \changesII{We also show
(in Section 6.4 of our PACT 2015 paper~\cite{medic})}
that our mechanism implicitly takes reuse information 
into account.}
%and that it correctly applies bypassing decisions
%appropriately based on warp type.
%the reuse based mechanism can be hard to
%implement at the shared L2 cache; 
%however, we use the Bloom filter based
%mechanism proposed for the Evicted Address Filter cache~\cite{eaf-vivek} to
%simplify the process of tracking reuse information.

\textbf{Software-based Cache Bypassing.} \green{Concurrent work by Li et
al.~\cite{li-sc15} proposes a compiler-based technique that performs cache
bypassing using a method similar to PCAL~\cite{donglihpca15}. Xie et
al.~\cite{xie-iccad13} propose a mechanism that allows the compiler to perform
cache bypassing for global load instructions. Both of these mechanisms are different from
\titleShort{} in that \titleShort{} applies bypassing to \emph{all} loads and stores that
utilize the shared cache, without requiring additional characterization at the
compiler level.} Mekkat et al.~\cite{mekkat-pact13} propose a bypassing mechanism for when
a CPU and a GPU share the last level cache. Their key idea is to
bypass GPU cache accesses when CPU applications are cache sensitive, which is
not applicable to GPU-only execution.

\textbf{CPU Cache Bypassing.} 
%In addition to GPU cache bypassing, there is prior work that proposes cache
%bypassing as a method of improving system performance for
%CPUs~\cite{tyson-micro28,doung-micro45,gaur-isca11,piquet-acsac07,teresa-micro97,teresa-isca97}. 
There are also several other CPU-based cache bypassing techniques. These
techniques include using additional buffers track cache statistics to predict 
cache blocks that have high utility based on reuse count
~\cite{annex-cache,gaur-isca11,zhang-ispled14,kharbutli-ieeetran,etsion-tc,chaudhuri-pact12,xiang-ics09,liu-micro08}, 
reuse distance~\cite{doung-micro12,gupta-ipdps13,chaudhuri-pact12,feng-interact12,park-sc13,yu-dasc,gao2010dueling,youfeng-micro02},
behavior of the cache block~\cite{jalminger-iccp03} 
or miss rate~\cite{mct,tyson-micro95,dbi,memik-hpca03}
As they do not operate on SIMD systems, these mechanisms do not (need to)
account for memory divergence heterogeneity when performing bypassing decisions.
%These mechanisms are only applicable to their
%targeted systems, and cannot be ported for GPGPU applications. 

\textbf{Cache Insertion and Replacement Policies.} Many works 
propose different insertion policies for CPU
 systems (e.g.,~\cite{bip,eaf-vivek,dip,rrip,khan-hpca14,mlp-aware}).  We compare our insertion policy against
the Evicted-Address Filter (EAF)~\cite{eaf-vivek}, and show
that our policy, which takes advantage of inter-warp divergence
heterogeneity, outperforms EAF. Dynamic Insertion Policy
(DIP)~\cite{dip} and Dynamic Re-Reference Interval Prediction
(DRRIP)~\cite{rrip} are insertion policies that account for cache
thrashing. The downside of these two policies is that they are unable to
distinguish between high-reuse and low-reuse blocks in the same
thread~\cite{eaf-vivek}. The Bi-modal Insertion Policy~\cite{bip}
dynamically characterizes the cache blocks being inserted. None of
these works take warp type 
characteristics or memory divergence behavior into account.
\changesII{Other work proposed prefetch-aware insertion and replacement 
policies~\cite{ebrahimi-isca2011,seshadri-taco2015,fdp}. \titleShort can be combined with such policies.}

%our insertion mechanism will also impact the
%replacement policy, and that we track cache block characteristics at a warp-level
%granularity. 
%The replacement policy in \titleShort{} also replaces blocks based
%on the hit ratio (where the mosly-miss category will be the prime candidate for
%replacement). 
%All these previously proposed policies does not take into
%account of the inter-warp heterogeneity in GPGPU applications.

%HERE - RACHATA

\textbf{Memory Scheduling.} 
Yuan et al. propose a GPU interconnect design that rearrange the sequence of memory
requests that arrive at each memory channel to reduce the complexity of GPU memory
scheduler~\cite{complexity}.
Chatterjee et al. propose a GPU memory
scheduler that allows requests from the same warp to be grouped
together, in order to reduce the memory divergence across different memory
requests within the same warp~\cite{chatterjee-sc14}. Jog et al. propose a GPU memory scheduler
that exploit the criticality information of warps in the GPU cores in order to
improve the performance of GPGPU applications~\cite{adwait-critical-memsched}.
\changesII{Principles of \titleShort can be incorporated into these schedulers.}
%Our memory scheduling
%mechanism is orthogonal to these approaches because we aim to reduce the interference
%that mostly-hit warps, which are sensitive to high memory latency, 
%experience due to mostly-miss warps.  
%It is possible to combine these two
%scheduling algorithms, by batching requests (the key mechanism
%from Chatterjee et al.~\cite{chatterjee-sc14}) for both the high and low priority queues (the
%key mechanism of our memory scheduler). 

% and as shown
%by Ausavarungnirun et al., these CPU schedulers are not applicable to GPU
%applications~\cite{sms}. Additionally, the design in Ausavarungnirun et
%al.~\cite{sms} is not applicable to GPU-only applications because the scheduler
%will fall back to FR-FCFS~\cite{fr-fcfs} when running on GPU-only applications.

%\noindent\textbf{Thread Throttling:} \yellow{Previous works propose thread
%throttling as another alternative solution to cache thrashing
%problem~\cite{ccws, ccws-swl,nmnl-pact13,cpugpu-micro,kuo-throttling14}.
%Combining \titleShort with such throttling mechanisms is an interesting
%direction for future work.}

There are several memory scheduler designs
that target systems with
CPUs~\cite{fr-fcfs,parbs,stfm,atlas,tcm,bliss,mise,pam,ipek-isca08,morse-hpca12,pa-micro08,cjlee-micro09,vwq-isca10,lee2010dram,mutlu-podc08,ghose2013,xiong-taco16,liu-ipccc16,bliss-tpds,lavanya-asm,memattack,mcp,jishen-firm,hyoseung-rtas14,hyoseung-rts16},
and heterogeneous compute elements~\cite{sms,jeong2012qos,dash-taco16}.  Memory
schedulers for CPUs and heterogeneous systems generally aim to reduce
interference across different applications.

\textbf{Improving DRAM.}
An alternative approach to mitigate memory divergence is to improve the performance
of the main memory itself. Previous works propose new DRAM designs that are capable of reducing memory latency in conventional
DRAM~\cite{chang-sigmetric16,lisa,dsarp,al-dram,tl-dram,ava-dram,donghyuk-stack,salp,donghyuk-ddma,micron-rldram3,sato-vlsic1998,hart-compcon1994,hidaka-ieeemicro90,hsu-isca1993,kedem-1997,son-isca2013,luo-dsn2014,chatterjee-micro2012,phadke-date2011,shin-hpca2014,chandrasekar-date2014,o-isca2014,zheng-micro2008,ware-iccd2006,ahn-cal2009,ahn-taco2012,superfri,imw2013,chang-sigmetric17,chargecache,softmc,lisa,seshadri2013rowclone,ambit,hsieh-iccd2016,loh-stack,kim-cal2015,dsarp,hashemi-micro2016,ahn-isca2015-2,ahn-isca2015}
as well as non-volatile
memory~\cite{meza-weed2013,ku-ispass2013,meza-cal2012,yoon-iccd2012,qureshi-isca2009,qureshi-micro2009,lee-isca2009,lee-ieeemicro2010,lee-cacm2010,ubm,banshee,youyou-iccd14,thynvm}.
Data compression techniques can increase the effective DRAM
bandwidth~\cite{bdi-pact12,lcp-micro13,toggle-hpca16,compress-reuse-hpca15,caba}.
All these techniques are orthogonal to \titleShort and can be used to further improve
the performance of GPGPU applications.

\textbf{Other Ways to Handle Memory Divergence.} \green{In addition to cache
bypassing, cache insertion policy, and memory scheduling, other works
propose different methods of decreasing memory
divergence~\cite{ccws,nmnl-pact13,cpugpu-micro,kuo-throttling14,largewarps,warpsub,zheng-cal,caws-pact14,tor-micro13}.
These methods range from thread throttling~\cite{ccws,nmnl-pact13,cpugpu-micro,kuo-throttling14} to warp
scheduling~\cite{ccws,largewarps,warpsub,zheng-cal,caws-pact14,tor-micro13}.
While these methods share our goal of reducing memory divergence, none of them
exploit inter-warp heterogeneity and, as a result, are \changesII{either} orthogonal or
complementary to our proposal.  Our work also makes new observations about 
memory divergence not covered by these works.}
% While these methods share our goal of reducing memory divergence, their
% solutions are orthogonal or complementary to our proposal and thus can be combined with our techniques.
% \yellow{Furthermore, we contribute new observations about memory divergence not
% covered by these works.}}

\section{Potential Impact}

While the problem that \titleShort is trying to solve, which is memory divergence,
is not new, key findings in this work provide novelty and create potential
research topics for the future. \changesIII{We discuss at least three such opportunities
and future directions.}

\textbf{Taking Advantage of Memory Divergence Heterogeneity.} 
\titleShort modifies the
memory hierarchy to introduce awareness of the memory divergence heterogeneity
between different types of warps. 
There are many other applications that can exploit warp type information.
Other resources within the GPU (e.g., GPU cores, warp scheduler) can exploit the memory divergence heterogeneity across
different warps to further improve the performance of GPGPU applications. 
For example, the warp type information can be used by the warp scheduler and thread block scheduler to
ensure that they do not schedule warps \changesIII{of the same type} to execute at the same time,
to limit the amount of cache contention that occurs.  
% Our
% key observation on memory divergence heterogeneity along with the warp-type
% identification logic can be used to assist multiple components within GPUs
% For example, warp-type information can be used by warp scheduler and thread block scheduler. 
Incorporating
the warp type information with other techniques, such as assist warps to relieve execution bottlenecks~\cite{caba}, can
enable GPUs to utilize resources based on the type of warps the GPU is executing.
\changesIII{For example}, mostly-hit warps favor a mechanism that provides low memory latency, while mostly-miss 
warps \changesIII{might} favor a mechanism that provides higher off-chip bandwidth.
Memory divergence heterogeneity can also be used to assist \changesIII{GPU} resource virtualization~\cite{zorua},
as virtual resource allocation can take into account the utilization of shared memory resources 
to determine how much of a particular memory resource to allocate to each thread block.

\changesIII{Warp type information can be used to improve the performance
of GPU address translation. 
Prior works~\cite{mask,mosaic} show that address translations that do not hit
in a TLB can incur long-latency page table walks, which can affect hundreds of
application threads at once.  Such long-latency address translations 
might have a greater impact on warps that are latency sensitive (e.g., \emph{mostly-hit} and \emph{all-hit}
warps).  Thus, warp-type information can be combined with previously-proposed
techniques that aim to reduce the overhead of GPU address translation~\cite{mask,mosaic}
to provide synergistic performance benefits.}

\changesIII{We believe the idea of warp-type heterogeneity enables many
different mechanisms to customize execution on a GPU to achieve higher
performance and energy efficiency. Hence, our PACT 2015 paper~\cite{medic}
paves the way for fine-grained customization of a GPU.}

%In terms of future impact, 
\textbf{Identifying Long-Latency Threads in a Warp.}
%\titleShort expands GPU research on two fronts. The
%first is how to 
Our PACT 2015 paper~\cite{medic} shows how to intelligently reduce the memory latency of threads within
a warp in order to \changesIII{reduce} the memory divergence problem. 
%Because GPU on-chip caches only provide latency 
%benefit for all-hit warps, caching in GPUs has been primarily used to the filter out
%requests and lower the off-chip bandwidth of GPGPU applications.  In this work,
%we provide a mechanism that selectively allows other warp types to benefit from
%the low latency of cache hits. 
%%Our observation on GPU's memory divergence 
%%heterogeneity exposes other research topics that can be used to accelerate long-latency 
%%threads for mostly-hit warps. Because 
However, \titleShort focuses on reducing the stall time of mostly-hit warps.
Long-latency threads \changesIII{can still exist in the mostly-hit warps} due to other
problems such as load balancing at the memory partitions. Additional work on 
(1)~how to identify \changesIII{latency-critical} threads within a warp and (2)~how to accelerate
these specific threads can further improve the performance \changesIII{and energy efficiency} of GPGPU
applications.
%In addition, providing a memory and cache partitioning scheme to reduce memory divergence
%can help mitigating memory divergence even further.

\textbf{Reducing \changesIII{High Queuing Delays and Memory Contention} in the GPU Memory Hierarchy.} 
As shown in our PACT 2015 paper~\cite{medic}, 
the queuing delay of throughput processors such as GPUs can become a performance bottleneck,
as the delay increases the stall time of warps of \emph{all} types. While the proposed
warp-type-aware cache bypassing mechanism in \titleShort aims to reduce the
queuing delay, non-uniform memory access \changesIII{patterns} can still cause contention at
a few L2 cache banks and memory partitions. In future systems, the parallelism
of throughput \changesIII{processors is likely to increase} further.  For example, future GPUs
will likely come with a higher number of GPU cores and \changesIII{larger SIMD widths.}~\changesIII{This} is 
expected to greatly increase the amount of contention and, thus, queuing delay,
for many resources.  The different components of \titleShort can serve as a
starting point \changesIII{for} future research on alleviating \changesIII{cache and memory} contention in future systems,
and can ultimately enable a larger amount of thread-level parallelism. \changesIII{We believe studying
the mitigation of high cache and memory contention is very promising for future parallel throughput
processors and encourage future work in this area.}
% research that
% provides radical changes in GPU caching might be required to be able to handle
% the parallelism of the GPU cores.

\section{Conclusion}

Warps from GPGPU applications exhibit heterogeneity in their memory divergence
behavior at the shared L2 cache within the GPU.  We find that (1)~some warps 
benefit significantly from the cache, while others make poor use of it;  
(2)~such divergence behavior for a warp tends to remain stable for long periods of the
warp's execution; and (3)~the impact of memory divergence can be
amplified by the high queuing latencies at the L2 cache.

%We have characterized the memory divergence heterogeneity amongst warps within
%a GPGPU application. The degree of memory divergence is a characteristic of each warp, and
%this heterogeneity differentiate warps that benefit from caching from warps that
%do not benefit from caching. We found that the latency tolerance heterogeneity 
%is also consistent and can be characterized. We also
%find that such heterogeneity can result in large, unnecessary queuing delays at
%the shared cache. 

We propose \emph{\titleLong} (\titleShort{}), whose key idea is to
identify memory divergence heterogeneity online in hardware and use this information to drive cache
management and memory scheduling, by prioritizing warps that take the greatest
advantage of the shared cache. To achieve this, \titleShort{} consists of three
\emph{warp-type-aware} components for (1)~cache bypassing, (2)~cache insertion,
and (3)~memory scheduling.  \titleShort{} delivers significant performance and
energy improvements over multiple previously proposed policies, and over a
state-of-the-art GPU cache management technique.  We conclude that exploiting inter-warp
heterogeneity is effective, and hope future works explore other ways of
improving systems based on this key observation \changesII{of our work}.

\section*{Acknowledgments}

We thank the anonymous reviewers and SAFARI group
members for their feedback. Special thanks to Mattan Erez
for his valuable feedback on our PACT 2015 paper. We acknowledge the support
of our industrial partners: Facebook, Google, IBM, Intel,
Microsoft, NVIDIA, Qualcomm, VMware, and Samsung.
This research was partially supported by the NSF (grants
0953246, 1065112, 1205618, 1212962, 1213052, 1302225,
1302557, 1317560, 1320478, 1320531, 1409095, 1409723,
1423172, 1439021, and 1439057), the Intel Science and
Technology Center for Cloud Computing, and
the Semiconductor Research Corporation.

{
\bibliographystyle{IEEEtranS}
\bibliography{references}
}

\end{document}